# Experimental Determination of the Structural Coefficient of Restitution of a Bouncing Asteroid Lander

Jens Biele[(1)], Lars Kesseler[(1,2)], Christian D. Grimm[(3)], Silvio Schröder[(3)], Olaf Mierheim[(3)], Michael Lange[(4)], and Tra-Mi Ho[(3)]

[(1)] DLR – German Aerospace Center, Micro-Gravity User Support Center, Cologne, Germany,
e-mail: Jens.Biele@dlr.de
[(2)] Fachhochschule Aachen, 52066 Aachen, Germany
[(3)] DLR – German Aerospace Center, Institute of Space Systems, Bremen, Germany
[(4)] DLR – German Aerospace Center, Institute of Composite Structures and Adaptive Braunschweig, Germany

The structural coefficient of restitution describes the kinetic energy dissipation upon low-velocity (~0.1 m/s) impact of a small asteroid lander, MASCOT, against a hard, ideally elastic plane surface. It is a crucial worst-case input for mission analysis for landing MACOT on a 1km asteroid in 2018. We conducted pendulum tests and describe their analysis and the results.

## Nomenclature

$A$ = amplitude of oscillation

$g$ = Earth gravity

$J$ = inertia, kg · m2

$CoG$ = Centre of Gravity, centre of mass

$LM$ = Landing Module

$MASCOT$ = Mobile Asteroid Surface Scout

Subscripts

$0$ = just before the collision

$f$ = just after the collision (restitution finished)

## I. Introduction

MASCOT, a small (10kg) Lander for small bodies in the Solar system, has been developed by DLR with important contributions by CNES during 2011-2014 for the Japanese Hayabusa-2 (H-2) space mission [1-3].

The target asteroid for this ambitious sample-return mission is asteroid 162173 Ryugu (provisional designation 1999 JU3), a primitive (C-type) Near-Earth asteroid of about 1 km diameter. H-2 has been launched Dec 4, 2014 and will reach its target in July 2018; Earth return is foreseen for 2020. The mothership will deploy MASCOT from an altitude of 50-100 with a separation speed of 5 cm/s, in a 15deg down angle. Impact velocity is estimated to be 15-19 cm/s depending on the actual mass and dimensions of the asteroid; this is less than 50% of the escape velocity. There is neither attitude control nor dedicated impact damping, such that the Lander will bounce uncontrolled until it finally comes to rest. A built-in actuator will be used for active relocation and uprightening of the Lander.

MASCOT mission analysis has the difficult task to predict, in a statistical fashion, the bouncing motion (map of final rest position, bouncing duration). The difficulty lies in the mechanical interaction between the MASCOT structure and the surface of the asteroid; it is modelled using the concept of a coefficient of restitution (COR), which a priori was unknown.

In this paper we describe the experimental determination of the (worst-case) COR of the MASCOT structure impacting a hard surface with representative velocities and attitudes. To our best knowledge, this is the first determination of coefficient of restitution for a spacecraft designed to soft-land.

In parallel, we are studying the complementary case of interaction of a rigid MASCOT body with a granular bed under micro-g conditions with soft-sphere discrete-element methods (D. Richardson, R. Ballouz, University of Maryland). This are computationally very expensive simulations with the full physical interaction (including rotation of the Lander) over a certain parameter space in impact velocity and angle, attitude, rotation state, particle size and size distribution. The results of of these investigations will be reported elsewhere.

## II. THE COEFFICIENT OF RESTITUTION

The process of rigid body collision is very complicated. The major characteristics are the very short duration and the large magnitude of the forces generated. Other phenomena include vibration waves propagating through the bodies, local deformations produced in the vicinity of the contact area, and frictional and plastic dissipation of mechanical energy. The COR simplifies the energy dissipation description during the complex plastic-elastic contact interaction of colliding solid bodies. Two different approaches [RD4 TN-1022] can be distinguished for impact and contact analysis: The impulse-momentum or discrete methods (confined primarily to impact between rigid bodies), and the continuous or force based analysis (which is more general). The coefficients of restitution etc. are the primary descriptive variables in the discrete methods, but can be derived only from the continuous method or by experiment (what we will try to do here).

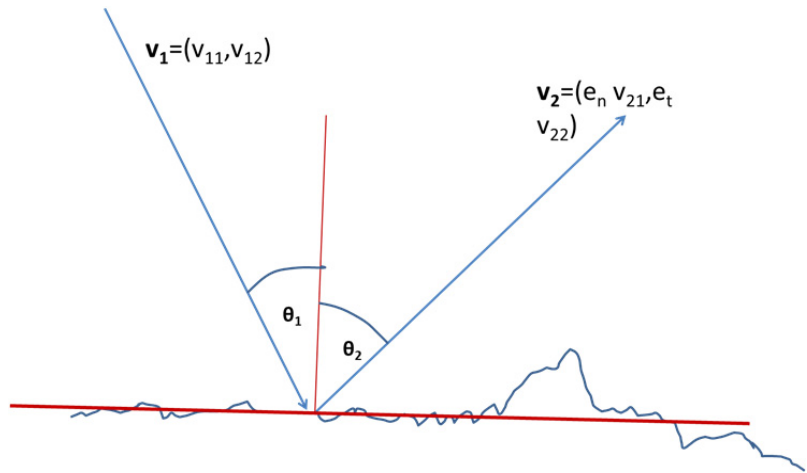

**Figure 1 Sketch of bouncing. Lander, treated as point mass, touches the surface at P (at angle $\theta_1$ to the surface normal n) with incoming velocity vector $v_1$ and bounces off at angle $\theta_2$ to n with outgoing velocity $v_2$. The surface normal refers to a size scale >> dimension of Lander. Surface roughness on scales of Lander and below is indicated by the blue "real" surface, it determines the probability distribution of $\theta_2$ and influences the rotation change of the real lander. $v_1$, n and $v_2$ do not necessarily need to lie in one plane.**

We use the following (Newton's) kinematic definition of the normal coefficient of restitution e:

$$e = e_n = -\frac{\mathbf{v}(t_f) \cdot \mathbf{n}}{\mathbf{v}(t_0) \cdot \mathbf{n}} = -\frac{v_{n,f}}{v_{n,0}} \quad (1)$$

where v is the vector of relative linear velocity at the contact point, $t_0$ is the start of the contact, $t_f$ the end of the contact. Thus, $e_n$ is defined as the ratio of the relative normal speed of separation to the relative normal speed of impact. There are more general definitions of $e$ which are discussed in [Stronge RD4]. In the general case of nonelastic collisions between two bodies, we can define a total COR $e = |\mathbf{v}_f|/|\mathbf{v}_0|$ and the energy loss $\Delta E$ is related to the coefficient of restitution by $\Delta E = \tfrac{1}{2}\mu v_0^2(1-e^2)$ where $\mu = m_1 m_2 / (m_1 + m_2)$ is the reduced mass and $v = |v_2 - v_1|$ is the relative speed between the two bodies. For perfectly elastic collisions $e = 1$; in the case of perfectly inelastic collisions, the two bodies stick together after the impact and $e = 0$.

A tangential coefficient of restitution can be defined, by analogy, as

$$e_t = \frac{v_{t,f}}{v_{t,0}} \quad (2)$$

To calculate the tangential coefficient of restitution, lateral friction is often applied according to Coulomb's law, $\mathbf{F_t} = -\mu F_n \hat{\mathbf{v}}_\mathbf{t}$.

The presently probably best "instantaneous" surface interaction model is that of [Tardivel 2014] which augments the COR with friction and rolling resistance. In particular, MASCOT is not a sphere but a cuboid; the response of a cuboid can be emulated, by choosing a suitable value of the coefficient of rolling resistance. Although the lander shape is modelled as spherical, the value of the coefficient of rolling resistance effectively enables a cube like behavior when the Lander is in lasting contact with the surface: it cannot easily roll. Although this model fails to capture the complex transfer of momentum (rotation/translation) that can happen for a cube during the first few bounces, it should be reliable for approximate statistic estimation and for highlighting deployment trade-offs.

### A. Taking into account rotation

In the literature, the COR is usually measured for impacting spheres, thus it is independent of attitude. In the general case the impact leads to a change of rotation of the impacting body which sensitively depends on the impact attitude, velocity, and the rotational state and moments of inertia matrix $\mathbf{J}$ of the Lander [Tardivel, 2013].

Figure 2 shows the possible partition of energy upon an impact. It is readily apparent that the real physical core of the COR is the energy dissipated into non-kinetic energy.

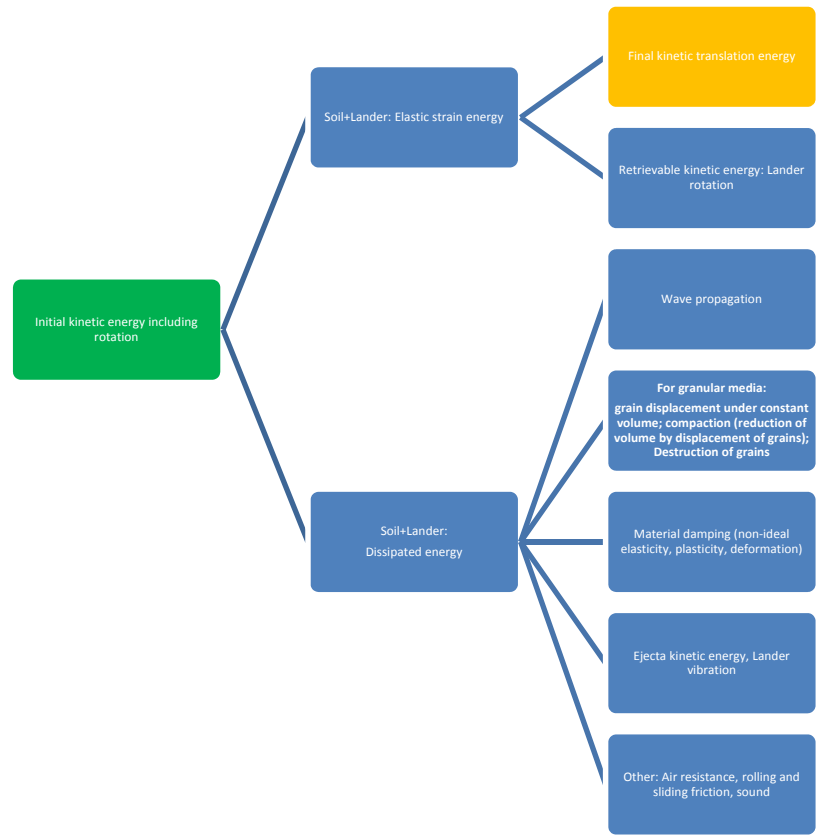

**Figure 2** Energy flow associated with an impact of an extended body on a surface

Excess kinetic energy (i.e. rotation and transverse linear motion), however, can be converted into normal linear motion and vice versa; for multiple bounces, this can lead to an apparent change of the effective COR at each bounce even if conditions are perfectly identical otherwise. What we measure here is the <u>"effective" coefficient of normal restitution:</u> this is the usual normal COR as defined above, including any rotational or non-perpendicularity effects. Only the normal velocity of the CoM counts. It could be >1 if enough rotational or transverse kinetic energy is converted into kinetic energy during the contact.

<u>"Structural" coefficient of restitution e\*:</u> COR freed of changes in rotational energy, energy in Y-translation and in the high-frequency mode of pendulum motion. We propose the following definition ($E_{kin} = E_{lin} + E_{rot} + E_{osc}$ is the total kinetic energy including rotation and oscillation):

$$e^* \equiv \sqrt{\frac{E_{kin,post}}{E_{kin,pre}}}$$

$$e^* = \sqrt{\frac{E_{lin,post} + \frac{1}{2}J\left[\omega_{post}^2 + \left(\hat{\varphi}_{hf,post}\omega_{hf,post}\right)^2\right]}{E_{lin,pre} + \frac{1}{2}J\left[\omega_{pre}^2 + \left(\hat{\varphi}_{hf,pre}\omega_{hf,pre}\right)^2\right]}}$$

$$E_{lin} = \frac{M}{2}(v_X^2 + v_Y^2) = \frac{M}{2}v^2 = \frac{M}{2}\left(\frac{v_X}{\cos\theta}\right)^2$$

$$J = \hat{\boldsymbol{\omega}}^T \mathbf{J} \hat{\boldsymbol{\omega}}$$

Thus, ideally we need, for the calculation of e*, not only the perpendicular velocities $v_X$, but also the incidence and emergent angles θ, the mass M and inertia tensor **J**, the direction (unit vector) of the vertical rotation axis $\hat{\boldsymbol{\omega}}$, the rotation rates $\omega$ around the vertical axis and the angular amplitudes and frequencies of the high-frequency pendulum mode.

**B. Velocity dependence**

Experimentally it is well known that the COR depends also on the impact velocity. Typically the COR decreases at higher speeds, attains a maximum <1 at a small velocity and goes to zero as the velocity approaches zero (sticking, due to e.g. van-der-Waals forces). According to theory of Thornton and Ning (1998); full analysis in Krijt[1] et al., (2012) for idealized spherical bodies the normalized curves shown in **Figure 3** are expected.

---

[1] Krijt, S., Tielens, A. G. G. M., Güttler, C., Heißelmann, D., and Dominik, C. (2012). Energy-dissipation channels in normal collisions. *Physical Review E*. submitted.

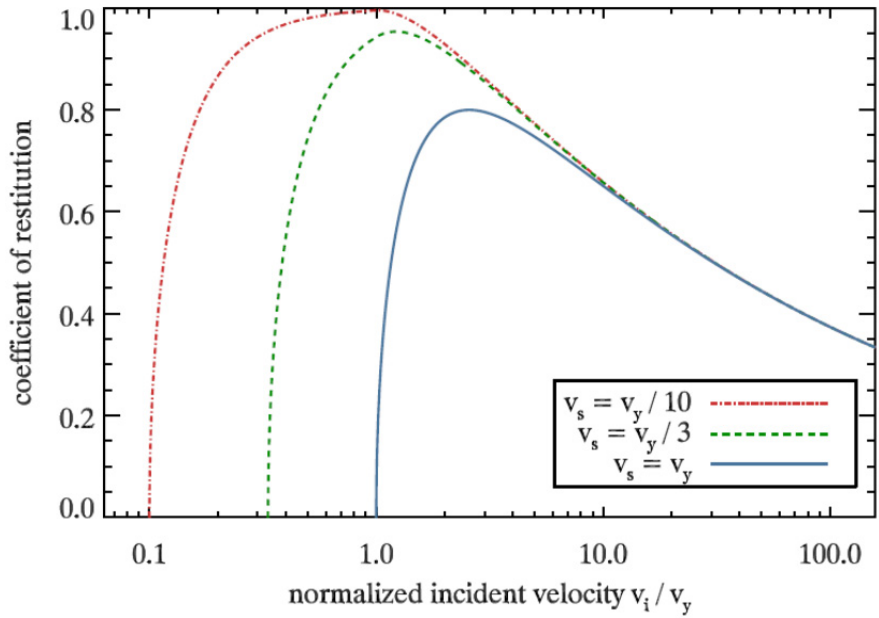

**Figure 3** Expected relation for the coefficient of restitution against the incident velocity (after Thornton and Ning, 1998 cited in Güttler et al., 2012)L $v_y$ is the so-called yield velocity

Hartmann et al. (1978) showed that catastrophic disruption of an impact target starts at velocities greater 2, 9, 37 m/s for dirt clods, ice and basalt, respectively. In the regime considered here, we are at much lower velocities such that no disruption nor even significant ejecta production is to be expected.

The coefficient of restitution of various rock samples on slabs of the same material is $e \approx 0.86 \pm 0.04$ for velocities < 1m/s (Imre, Räbsamen et al. 2008). The coefficient of restitution of 1 m diameter granite spheres is $e \approx 0.83 \pm 0.06$ for collisions between ~1-m-scale spheres at speeds of order 1 m s$^{-1}$. No clear trend of coefficient of restitution with impact speed was discernible in the data down to velocities of 0.14 m/s. (Durda et al., 2011, Imre 2008). Thus, by close analogy, we conclude that COR(concrete) is $e \approx 0.85 \pm 0.10$.

## C. Composite COR

For collisions between hard bodies that have dissimilar material properties, a composite coefficient of restitution (COR) can be calculated if, for each body, the material properties and COR values for self similar collisions are known. The composite COR represents the energy dissipated during a collision between materially-dissimilar bodies. This analysis of energy distribution during impact extends the usefulness of the measured values for COR by obtaining the separate dissipation from each body of a collision pair [Coaplen and Stronge, 2004].

$$e^{*2} = \frac{k_1 e_2^2 + k_2 e_1^2}{k_1 + k_2} \quad (1)$$

Where the k's are the compliances (linear rate-independent compliance, yield stress Y (generally better) or elastic module E) and the $e_i$'s the CORs of impacts at the work-equivalent velocity between bodies of the same material. This formula improves/corrects/replaces the formula by Hodgkinson (1834) which has often been referenced (e.g. Goldsmith, Impact, p. 7, eq. 2.14).

### III. TEST OBJECTIVE AND PLAN

The Impact Energy dissipation test shall show the dynamic behaviour of the LM when hitting a solid wall. The coefficient of restitution for ~10-20 cm/s impact velocities in "all possible orientations" is to be measured and imparted rotation is to be documented.

**Table 1 Test plan**

| Case No. | Run No | $V_0$ | attitude | L | LM did rotate strongly? |
|---|---|---|---|---|---|
| [-] | [-] | [m/s] | [-] | [m] | [Y/N] |
| 1 | 1 | 0.1 | +x | 5.150 | y |
|   | 2 | 0.1 | +x |   | y |
|   | 3 | 0.1 | +x |   | y |
| 2 | 1 | 0.1 | +y |   | y |
|   | 2 | 0.1 | +y |   | y |
|   | 3 | 0.1 | +y |   | y |
| 3 | 1 | 0.1 | +z | 5.150 | y |
|   | 2 | 0.1 | +z |   | y |
|   | 3 | 0.1 | +z |   | y |
| 6 | 1 | 0.1 | -z | 5.150 | y |
|   | 2 | 0.1 | -z |   | y |
|   | 3 | 0.1 | -z |   | y |
| 9 | 1 | 0.2 | corner x-y-z | 5.080 | n |
|   | 2 | 0.2 | corner x-y-z |   | n |
|   | 3 | 0.2 | corner x-y-z |   | n |
| 10 | 1 | 0.2 | corner -xyz | 5.080 | ? |
| | 2 | 0.2 | corner -xyz | | ? |
| | 3 | 0.2 | corner -xyz | | ? |
| 13 | 1 | 0.2 | +z | 5.150 | y |
| | 2 | 0.2 | +z | | y |
| | 3 | 0.2 | +z | | y |
| 16 | 1 | 0.2 | -z | 5.150 | y |
| | 2 | 0.2 | -z | | y |
| | 3 | 0.2 | -z | | y |
| 19 | 1 | 0.2 | corner x-y-z | 5.080 | n |
| | 2 | 0.2 | corner x-y-z | | n |
| | 3 | 0.2 | corner x-y-z | | n |
| 20 | 1 | 0.2 | corner -xyz | 5.080 | n |
| | 2 | 0.2 | corner -xyz | | n |
| | 3 | 0.2 | corner -xyz | | n |
| 23 | 1 | 0.1 | edge -y-z | 5.080 | n |
| | 2 | 0.1 | edge -y-z | | n |
| | 3 | 0.1 | edge -y-z | | n |
| 24 | 1 | 0.2 | edge -y-z | 5.080 | n |
| | 2 | 0.2 | edge -y-z | | n |
| | 3 | 0.2 | edge -y-z | | n |
| 25 | 1 | 0.1 | edge -x-y | 5.120 | y |
| | 2 | 0.1 | edge -x-y | | y |
| | 3 | 0.1 | edge -x-y | | n |
| 26 | 1 | 0.2 | edge -x-y | 5.120 | y |
| | 2 | 0.2 | edge -x-y | | y |
| | 3 | 0.2 | edge -x-y | | y |
| 27 | 1 | 0.1 | -x bounce knob | 5.100 | n |
| | 2 | 0.1 | -x bounce knob | | n |
| | 3 | 0.1 | -x bounce knob | | n |
| 28 | 1 | 0.2 | -x bounce knob | 5.100 | y |
| | 2 | 0.2 | -x bounce knob | | y |
| | 3 | 0.2 | -x bounce knob | | n |
| | 4 | 0.2 | -x bounce knob | | y |

The movement is to be documented by cameras specifically from the side and below. The motion and absolute speed independent of parallax errors is also measured with a laser distance measurement. The impact velocity is set by choosing the initial deflection of the pendulum in the rest position.

## IV. THE TEST SETUP

### D. Test article

The impact energy dissipation test was performed on the MASCOT LM STM 2.2 (Structural and thermal test model, henceforth abbreviated with LM) which is equipped with similar mass dummies as used for previous drop tower campaigns. With the help of trim masses the mass and CoG of the LM was brought close to the FM values. Moments of inertia have not been measured on the LM; the (CAD calculated) FM values are used instead.

The geometrical and mass properties of the LM are given in Table 2 and Table 3.

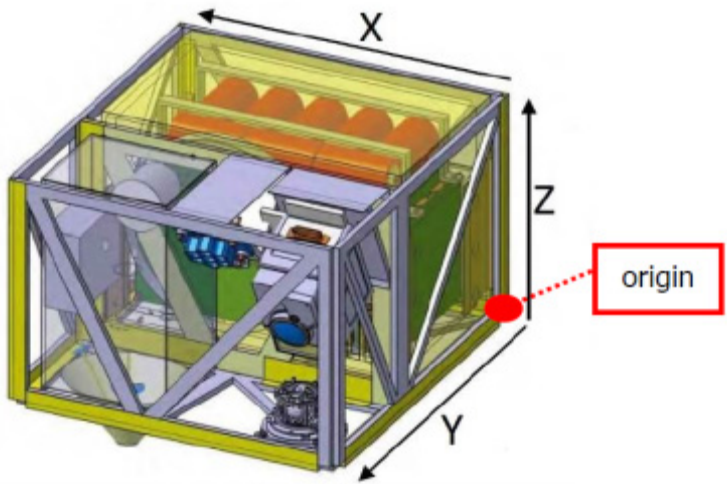

**Figure 4 MASCOT CAD drawing with definition of body-fixed coordinate system**

The laboratory reference system is defined as follows: +X in the horizontal direction of the main pendulum swing at rest position; +Z up, parallel to the vertical rope at rest position; +Y completes the right-handed triad (i.e., horizontal, away from the HS1 camera). The origin of this XYZ system is arbitrary.

**Table 2 Geometrical properties of LM**

| Dimensions | (mm) |
| --- | --- |
| Length (y) | 292.4 |
| Width (x) | 277.4 |
| Height (z) | 197.3 (191.3 CFC frame only) |

**Table 3 Mass properties of LM**

| Mass (kg) | 9.403 | | |
|---|---|---|---|
| Center of Mass | Test article | Flight model | Delta |
| CoM x (mm) | 145±1 | 149 | 0±1 |
| CoM y (mm) | 140±1 | 142 | 4±1 |
| CoM z (mm) | 103±1 | 103 | 2±1 |
| Inertia tensor with respect to CoM (kgm²), axes parallel to coordinate axes (kgm²), FM | $I_{CoM} = \begin{pmatrix} I_{xx} & I_{xy} & I_{xz} \\ I_{yx} & I_{yy} & I_{yz} \\ I_{zx} & I_{zy} & I_{zz} \end{pmatrix} = \begin{pmatrix} 0.099 & -0.0051 & 0.0019 \\ -0.0051 & 0.0825 & 0.0005 \\ 0.0019 & 0.0005 & 0.121 \end{pmatrix}$ | | |

### E. Pendulum setup

The LM hangs from a support by a rope of ~5m in front of a thick concrete wall. The LM is supported in its chosen attitude by 3 ropes distributed around the CoG. The test article thus has exactly the 3 degrees of freedom we require: translation both in X- and Y-directions (fundamental pendulum mode) and rotation around the vertical rope axis (the effective shear module of the rope and thus the restoring torsional momentum is very small, thus the rotation around Z is practically unhindered).

The rope had a Dyneema core (11 strands), diameter 2.0±0.2 mm and a high-strength braided polyester mantle, outer diameter 2.62±0.10 mm. Mass per length = 4.54..4.59 g/m (mantle 2.37±0.16g/m, core 2.18±0.16g/m). Stiffness was estimated as 50200 N/m.

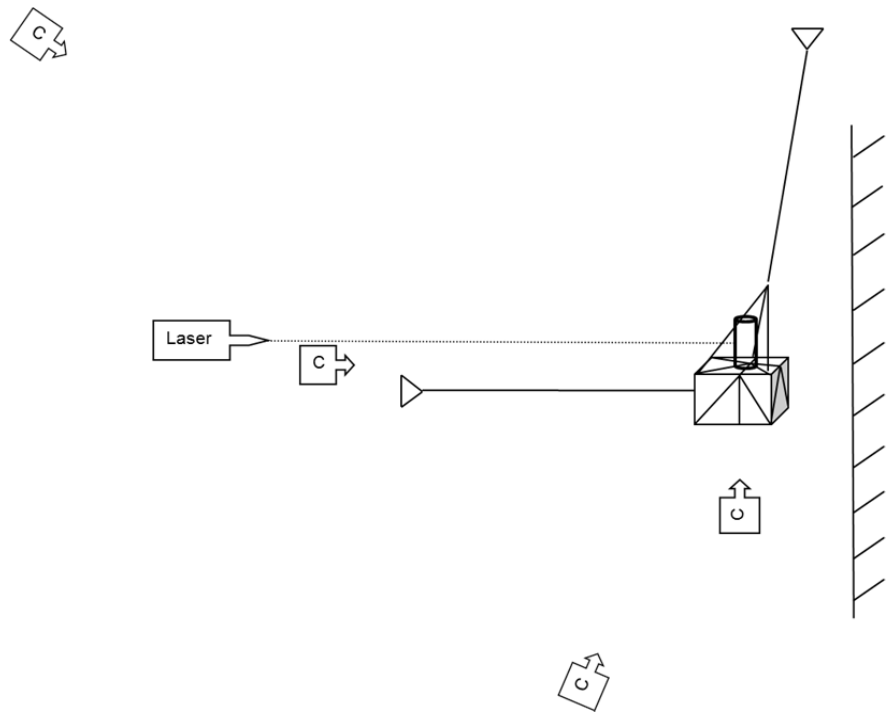

**Figure 5 Scheme of test setup**

The pivot of the pendulum is a steel hook at the end of the robotic arm (nominal static bearing capacity of 500 kg) of the Landing & Mobility Test Facility LAMA at ZARM. Three ~1.4 m long ropes attached to the LM and converging to a knot and the vertical portion of the rope were used to set the desired attitude. By remotely fine-controlling the robotic arm, the LM rest position could be precisely adjusted such that it almost (0-1 mm) touched the wall. The LM was then deflected to the desired initial pendulum amplitude (0.14 and 0.07 m, respectively, giving 0.2 and 0.1 m/s initial collision velocity) and released by opening a mechanical clamp holding the deflection rope. The total pendulum length (pivot to CoM of the LM) was roughly measured by another laser distance sensor, it is accurate to ~10cm. The pendulum/collision motion was tracked with 4 high-speed cameras (from behind, below, side, and from an isometric view) and with a laser distance sensor along the main pendulum plane (Figure 6). For a few test cases (#1, 3, 10, 25, 27) characterization measurements (laser data for a minute or so of free pendulum swing) were done, giving the precise period and the free damping constant, which was found δ=0.0025±0.0003 (from 0.0006 to 0.0033, no clear correlation with attitude).

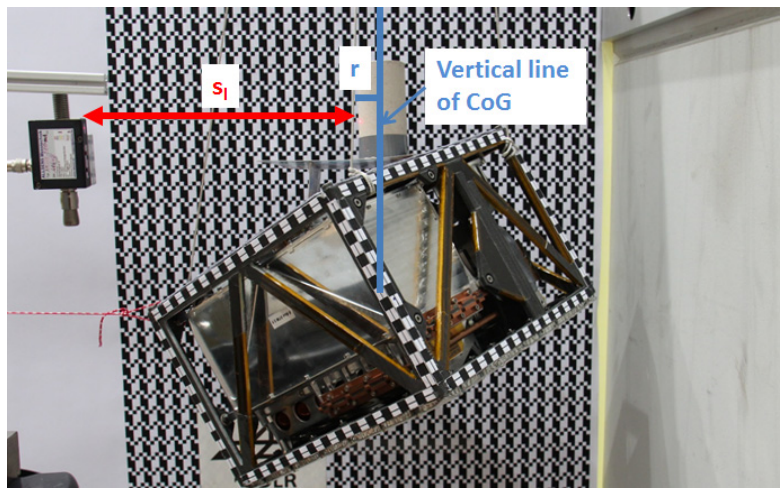

**Figure 6 Test article from the side in hold position with laser distance measurement shown (laser distance sl and cylindrical radius r=23mm)**

It turned out that the ideal case of impacting "flat to flat" with a LM face to the wall is difficult in practice: there is always a corner or edge contacting the wall first, so the impact to a face cannot be realized with the standard setup. A modification was therefore implemented for the last cases, where a face of the LM impacts about centrally a dedicated knob fixed on the wall (a brass weight glued to the concrete, see Figure 7).

We observe that these cases lead to the highest COR values; they correspond to MASCOT impacting a single roundish hard pebble in the center of a face, on an otherwise flat surface.

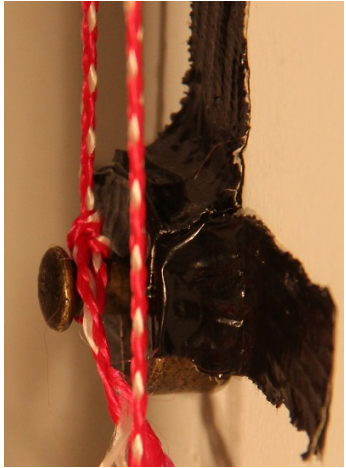

**Figure 7** Bounce knob, a standard brass weight (10g TBC), fixed to the wall with duct tape and a rope for strain-relief. Height over all 15.5±0.9 mm, diameter LM-side 6.1 mm, neck diameter 5.4 mm, base diameter 15.6 mm.

F. Sensor Equipment

| Identifier | Brand | Type | Frames per sec | Position | Image |
|---|---|---|---|---|---|
| HS1 | Photron | FASTCAM SA3 Model 120K 1024x1024 | 500 | 90° to X in 3-4 m distance | |
| HS2 | Casio | Exilim EX-F1 512x384 px TBC | 300 | Bottom ~10-20cm distance | |

| HS3 | Casio | Exilim EX-F1 512x384 px TBC | 300 | Back, in +X direction | 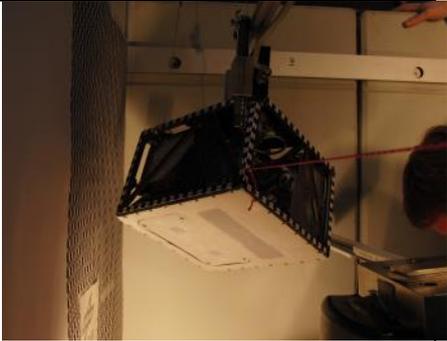 |
| ISO | Canon | EOS-650D 640 x 480 px EFS 18-55mm Objective | 25 | Diagonal view ("isometrical") | 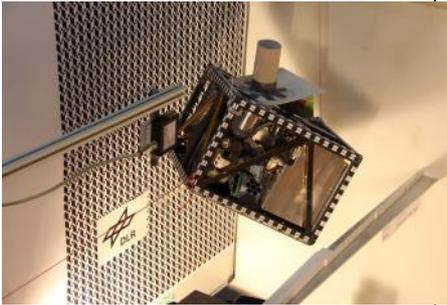 |

The laser distance sensor was a Allsens Meßtechnik Model AM300-500 Laser triangulation sensor, measurement range 125-625 mm, linearity ±0.40 mm, resolution 0.5 mm (10bit-discretization of output), sampling rate 600 Hz.

## V. TEST RESULTS AND OBSERVATIONS

All cameras worked as planned, only a few test runs the HS2 have been positioned wrong, so that the impact point is out of frame. The laser measures the distance of the LM in movement direction by pointing towards a board cylinder (diameter 46 mm, length 100 mm) positioned at the vertical CoG line (see Figure 6). The cylinder guaranties an accurate distance determination even if the LM is rotating after impact. On the other hand movements across the main movement direction lead to errors (max < R = 23 mm) due to the rounding of the cylinder. A sidewise motion can be counterchecked with the bottom or side camera (HS2/HS3). The bottom camera can be used for the determination of rotation around the vertical axis. For the analysis the LM and the background (front view) is covered with chessboard pattern of 5 and 10 mm size. In addition a ruler is placed directly beneath the front part of MASCOT to allow 1 mm precise analysis with HS1. For error correction the parallaxes have to be taken into account. It can be calculated geometrically with the known distances of each object to each other.

Two kinds of raw data were acquired: 1. video data (.mov, .avi) of all cameras and 2. laser distance data (.xls). Inspection of the data shows that the laser distance data have the known resolution of 0.50 mm (corresponding to an rms of ±0.15 mm) and the HS1 data can be tracked to an rms of ±0.1 mm, indicating super-resolution (less than 1 px). The time base of the laser sensor and the cameras is taken as exact (for forming the COR only the stability over a few sec is important, while for comparison between laser and cameras a worst case relative accuracy of any on-chip quartz oscillator of 1E-3 is still at least one order of magnitude better than other error sources).

With the laser distance sensor, per case and run typically 5 bounces could be observed, with the camera data usually 2. This gives us the possibility to extend our test cases to smaller impact speeds, for COR down to nearly 0, for $COR^*$ typically down to 0.05 m/s ($2^{nd}$ bounce).

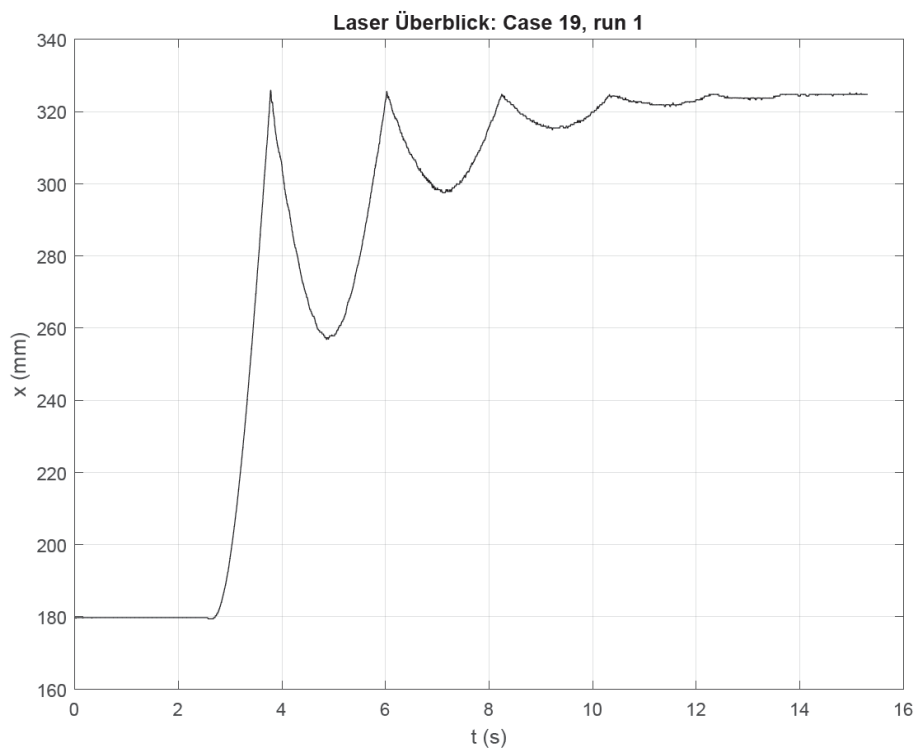

**Figure 8 A well-behaved example: Laser**

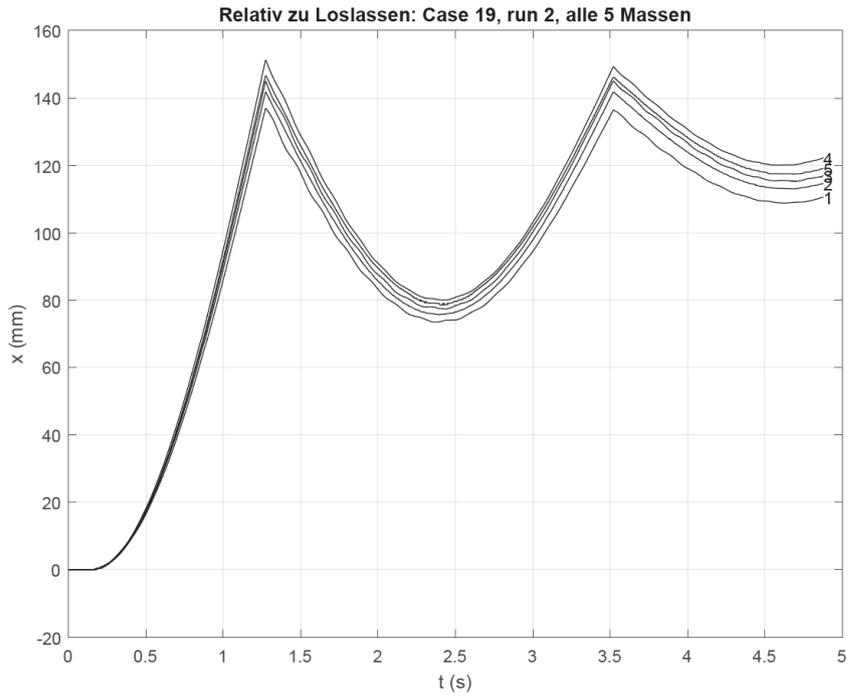

**Figure 9 A well-behaved example: HS1 with tracker**

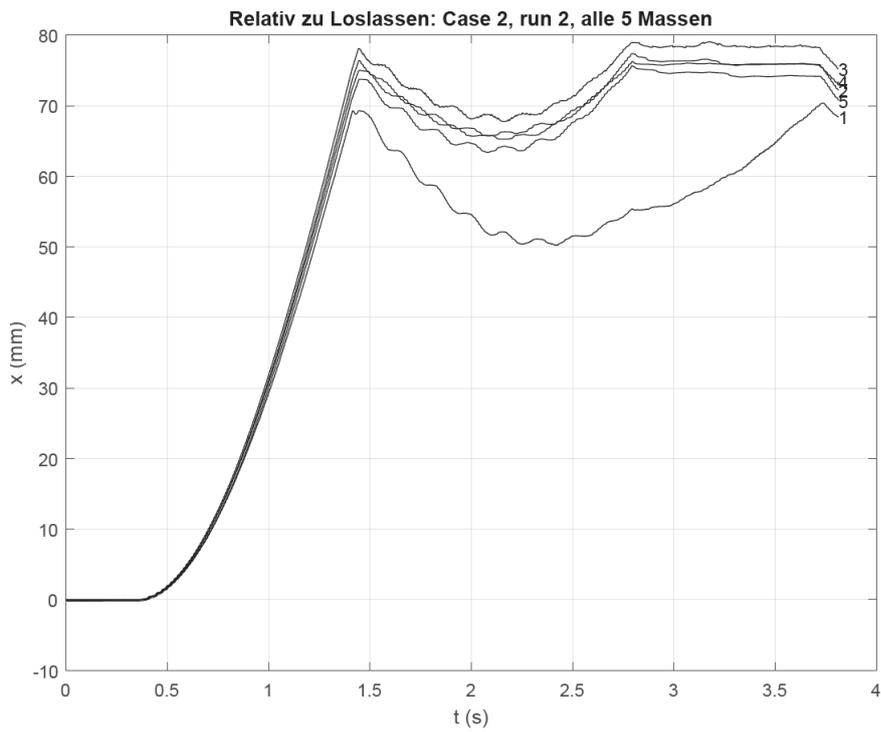

**Figure 10 A not so well-behaved example (HS1)**

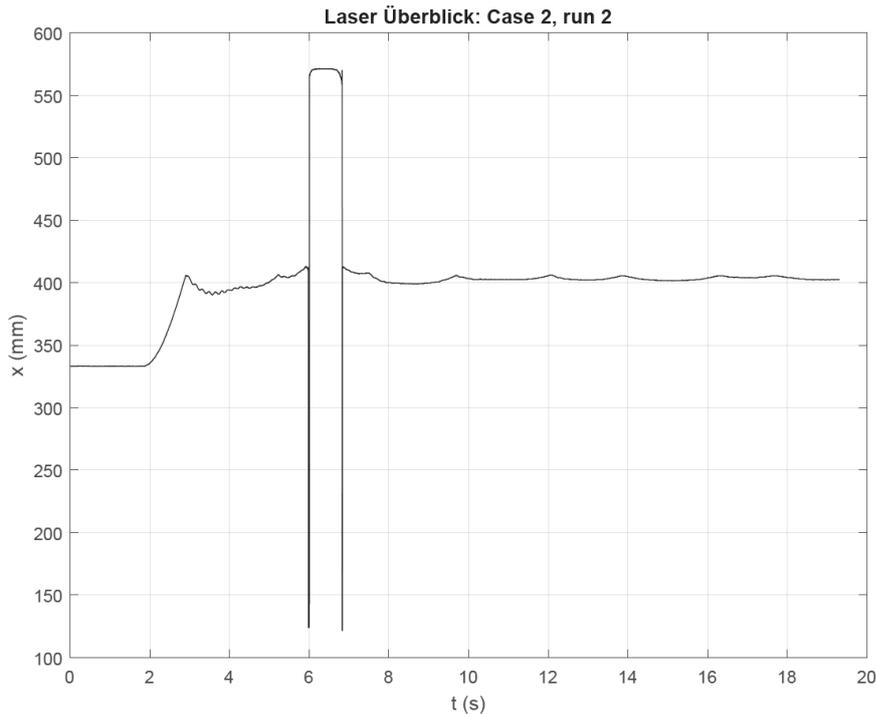

**Figure 11 A not so well-behaved example (Laser)**

A variety of bouncing behavior could be observed. While the test cases with an impacting corner or face-on-knob are mostly "well-behaved", impacts of face flat-on or impact of an edge often lead to considerable rotation after the first contact. In some cases the LM lingered for some seconds at the wall, consecutively rotating from one contacting edge to the other and back, until it finally bounced off the wall with a significant residual rotation. Some cases show a COR near 0 but a fast rotation after the bounce: here almost all translational energy has been converted into rotational energy. Visual inspection of the LM and the wall after the tests showed no obvious deformation or damage, just minute superficial changes at edges and corners of the LM and barely visible indentation in the wall.

## VI.   DATA ANALYSIS

The principle quantities we are interested in are the X- (and Y- for e*) components of the velocities of the CoM of the test article just before and just after each bounce (a bounce divides the curves in parts we will call segments henceforth) as well as the rotation rates around the vertical axis then. The data, on the other hand, give only distance of a certain point on the LM as a function of time. In the case of the laser distance sensor, this distance is modulated by the motion of the target cylinder and the 0.5 mm discretization but otherwise absolute except an unknown offset. The video camera data give actually the projection of points in the XZ-plane (on the swinging LM) into the camera focal plane (central projection, neglecting geometrical distortion in the optical system). Since camera HS1 is positioned at ~3 m to the LM, the projection can be regarded as parallel projection to first order. Even then, the tracking of the camera data does not immediately give the motion of the CoM, but the motion of the track points which is a combination of pendulum motion of the CoM, rotation and other effects. HS2 has been used to determine the rotation rate by comparison of two track points; here only the relative positions in a horizontal plane need to be known.

The main point for the analysis is that, to obtain velocities, naïve differentiation of the X(t) curves is bound to fail. There is far too much "noise" in the data; smoothing would be last resort, but introduces arbitrariness and new, unknown systematic errors. For the camera data, the rotation leads to systematic distortions in the tracked coordinates, such that any differentiation of the original data will introduce large systematic errors.

The solution lies in careful modelling the original data to physically justified fit functions for each segment. Once the numerous coefficients of these fits have been determined, the parameters of the simple spherical pendulum motion (amplitude, frequency, damping coefficient and phase) can be extracted and used to calculate the COR by simple mathematical analysis. The fit functions are nonlinear in the parameters, thus, sufficiently accurate starting values have to be supplied to assure convergence – to this end we estimated some geometrical parameters, manually determined the bounce times and constrained the parameters wherever possible.

Note that the ratio of speeds is inherently more accurate than the absolute speeds. The latter depend on the estimated distance from the pendulum pivot to the given point (laser spot on the target cylinder for the laser data, chosen track point for the camera data) and has an absolute accuracy of ±5cm, i.e., the absolute speed is known within a systematic uncertainty of 1%.

## G. TOOLS, DATA MODELING

The camera data have been analysed using the freeware tool tracker [reftracker], which is slow but very flexible and accurate. Later, we also tried the manufacturer-supplied software PHOTRON Fastcam Analysis-Software (PFA) [refPFA]. The numerical processing and fits have been performed with a student version of MATLAB making extensive use of the nonlinear solvers lsqcurvefit and lsqnonlin.

## H. THEORY

In this chapter, we derive the necessary equations of motion applicable to our test case.

*1. Pendulum motion (circular pendulum)*

(Reduced) length *l*, acceleration of gravity *g*, angular deflection $\varphi$; the initial deflection for time t=0 is $\hat{\varphi}$. Then, with the usual small-amplitude approximation and including free damping (by air resistance ~v and friction in the rope and the pivot):

$$\ddot{\varphi} + 2\delta\dot{\varphi} + \frac{g}{l}\varphi = 0$$

$$\omega_0 = \sqrt{g/l}; \quad T_0 = 2\pi\sqrt{l/g}$$

$$\varphi(t) = \hat{\varphi}e^{-\delta t}\sin(\omega t + \psi), \quad \omega = \sqrt{\omega_0^2 - \delta^2}, \quad \omega_0 > \delta$$

$$\dot{\varphi}(t) = -\delta\hat{\varphi}e^{-\delta t}\sin(\omega t + \psi) + \hat{\varphi}e^{-\delta t}\omega\cos(\omega t + \psi)$$

The linear tangential velocity in X is (damping neglected) proportional to the pendulum amplitude, length and frequency:

$$v = \dot{\varphi}l = \hat{\varphi}l\omega_0\sin(\omega_0 t + \psi)$$

$$s = \hat{\varphi}l : \text{arc length}$$

$$v_{max} = \hat{\varphi}l\omega_0 = \hat{\varphi}\sqrt{gl} = \hat{s}\omega_0$$

$$\Rightarrow COR = \frac{\hat{s}_{post}}{\hat{s}_{pre}}$$

The measurement of maximum amplitudes before and after a bounce are thus sufficient to measure the COR.

g=9.8132 m/s² in Bremen [http://www.ptb.de/cartoweb3/SISproject.php].

Now the amplitudes we used are very small, at maximum 1.6 deg (0.144 m /5.1 m). Thus the difference between linear amplitude A (deflection in X-direction) to the arc length is negligible. The maximum, i.e. impact velocity is given by $v_0 \approx \sqrt{g/l} \cdot A \approx 1.387\frac{1}{s} \cdot A$ and the height (Z) difference along the arc is < 2.0 mm. The sine (or cosine) of the motion in X can be fitted with a polynomial in *t* of 6[th] degree to an accuracy of always 2μm. The phase $\psi$ is known for the ideal case; in reality, it can be fitted to include non-ideal initial conditions (e.g., a small impulse at release).

*2. High-frequency pendulum modes, spherical pendulum modes*

A pendulum on a flexible string has higher modes of oscillation [Hughes 1953]; first high-frequency solution is a pitch oscillation around a point a few mm below the LM CoM. The frequency $\omega_1$ is observed as ~ 39 rad/s

(6 Hz) in our setup which is consistent with theory. It is strongly damped ($\delta \sim 1/s$) and its amplitude is up to 0.008 rad, excited by the collisions with the wall but it is also seen in the first segment, both in the laser and camera measurements, excited by the release.

The pendulum may swing not only in a plane (on a circle), but may describe an ellipse (i.e. it moves on a sphere). For a small transverse component of motion, this may be described in the theory of the spherical pendulum as two independent harmonic oscillations with the same frequency.

The fit function for the measured motion of a point P at mean distance $\overline{L}_P$ to the pivot, reduced to the CoM, in X-direction is thus:

$$\begin{aligned} X(t) &= \overline{L}_P \arcsin\left(Ae^{-\delta t}\cos(\omega t + \psi)/\overline{L}_{CoM}\right) \\ &+ e^{-\delta_1 t}(a\sin\omega_1 t + b\cos\omega_1 t) \\ &+ C \end{aligned}$$

C is a zero-point offset, $\delta$ the damping constant ($\delta \ll 1$ s$^{-1}$), $A$ is the fictitious linear amplitude of the CoM for t=0, $\omega$ the (rather well known) angular frequency. $\psi$ describes the phase. $a, b, \omega_1, \delta_1$ describe the hf pendulum mode. The L values are uncritical; we adopt the mean values, $L_{CoM}$ =5087.8mm, $L_P$ = $L_{CoM}$ - 200mm.

The motion of the CoM in Y-direction is given by

$$Y(t) = D + \sin\theta \cdot X + k\sin(\omega t + \psi_3)$$

A point not in the CoM will also experience rotation effects; in the camera data, central projection effects may play a role.

3. *Effects on the laser distance measurement*

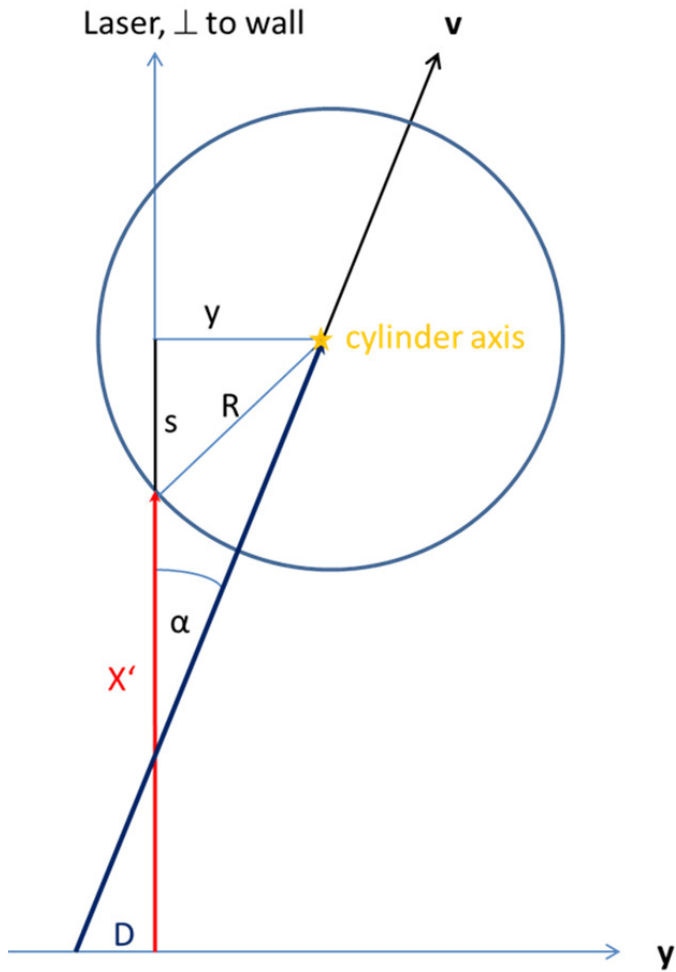

**Figure 12**

The net effect of lateral movement $y$ of the laser dot on the cylinder, caused by a deviation $\varepsilon=\sin\alpha$, offset D, periodic lateral terms by spherical pendulum mode (amplitude k, phase $\psi_3$) and rotation of the cylinder (rotation rate $\omega_d$ phase $\psi_2$) with an axis offset by r to the LM rotation axis is:

$$-\sqrt{R^2 - [X\varepsilon + D + r\sin(\omega_d t + \psi_2) + k\sin(\omega t + \psi_3)]^2} + r\cos(\omega_d t + \psi_2)$$

In some cases the starting values for the coefficients had to be found by Monte Carlo methods (in particular $\psi_2$ and $\psi_3$ proved very sensitive). It was also found that strong correlations exist between several fit coefficients making it mandatory to constrain them tightly between reasonable lower and upper limits.

4. *Effects of rotation and projection on camera data*

Rotation of the LM produces a cosine-term in the HS1-data, namely

$$+r_i \cos(\phi_i + w_D t)$$

where the $r_i$ are the horizontal distances of the tracked points from the rotation axis (assumed to be the vertical pendulum axis), the $\phi_i$ are their angular positions at the beginning of a segment (counted in the XY-plane from the +X direction) and $w_D$ is the rotation rate. The high-frequency pendulum mode can be written for the i=1..N track points (since the LM is a rigid body) as

$$+\exp(-\delta_1 t)(\Delta L_i + z_1)\left[a_1 \sin(w_1 t) + b_1 \cos(w_1 t)\right]$$

with the angular amplitude $\sqrt{a_1^2 + b_1^2}$ and a proportionality to the vertical distance $\Delta L_i + z_1$ from the center of oscillation ($z_1$ above the CoM which forms the reference for the pendulum length corrections $\Delta L_i$ of each track point).

For strictly accounting for central projection effects, one would need 6 additional parameters additional to the description of changing point coordinates by rotation. Here, we write approximatively for the measured X',

$$X' = \frac{X}{1 - \frac{r_i \cos(\varphi + w_D t)}{d}}$$

where the reference plane (in camera direction, Y) contains the vertical pendulum and rotation axis, $d$ can be set fixed and the $r_i$ absorb the B factor (~1) and $d$ uncertainty. This is a good approximation for $d >> X$ and corrects the most important error, namely that any motion in X has a different scale on the camera sensor depending on the distance of the point.

## 5. Other effects and non-idealities

Nonideality of the concrete wall

The massive concrete wall used in the tests is neither ideally elastic (e=1) nor infinitely rigid. We observed very minute indentations in the concrete after the tests. Here we estimate the corrections to be applied to get the COR of MSC against an either infinitely rigid (Y=∞) or ideal (e=1) wall of the same rigidity as concrete.
Using equation (1), with

$e_*$ measured

$e_1$ MSC-MSC unknown

$e_2$ concrete-concrete: ~0.83 assumed

$k_1$ (MSC) = 50 MPa (TN-1036, from FEM)

$k_2$(concrete)=30 GPa

$$e_1 = \sqrt{\frac{e_*^2(k_1+k_2) - k_1 e_2^2}{k_2}}$$

$e_1$(MSC-MSC) is always less than measured $e_*$; below $e_*$=0.034, e (true) is zero. At typical $e_*$, the difference is 1e-4 to 1e-3, rather negligible.

$e_{**}$ is MSC to a wall with infinite E: $e_{**}$=$e_1$. The e of concrete, $e_2$, does not not play any role then.

$e_{***}$ is MSC to a wall with finite E but $e_2$=1:

$e_{***} = \sqrt{\dfrac{k_1 + k_2 e_1^2}{k_1 + k_2}}$ is always greater than $e_*$. In the range 0.4-0.8, only by 3..6E-4.

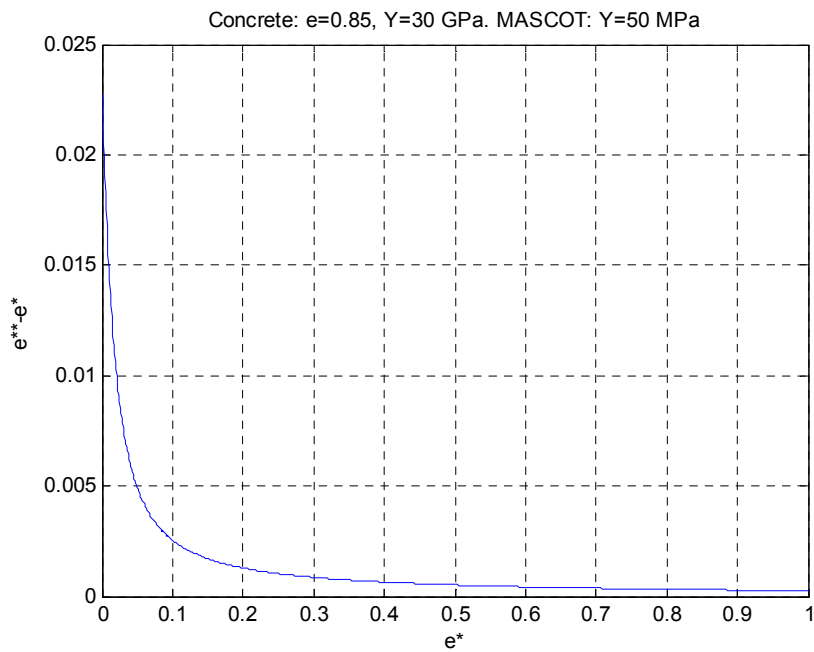

Figure 13 COR difference MSC-ideal hard surface (Y of concrete, but e=1) vs. measured

# VII. RESULTS

## I. Effective COR e

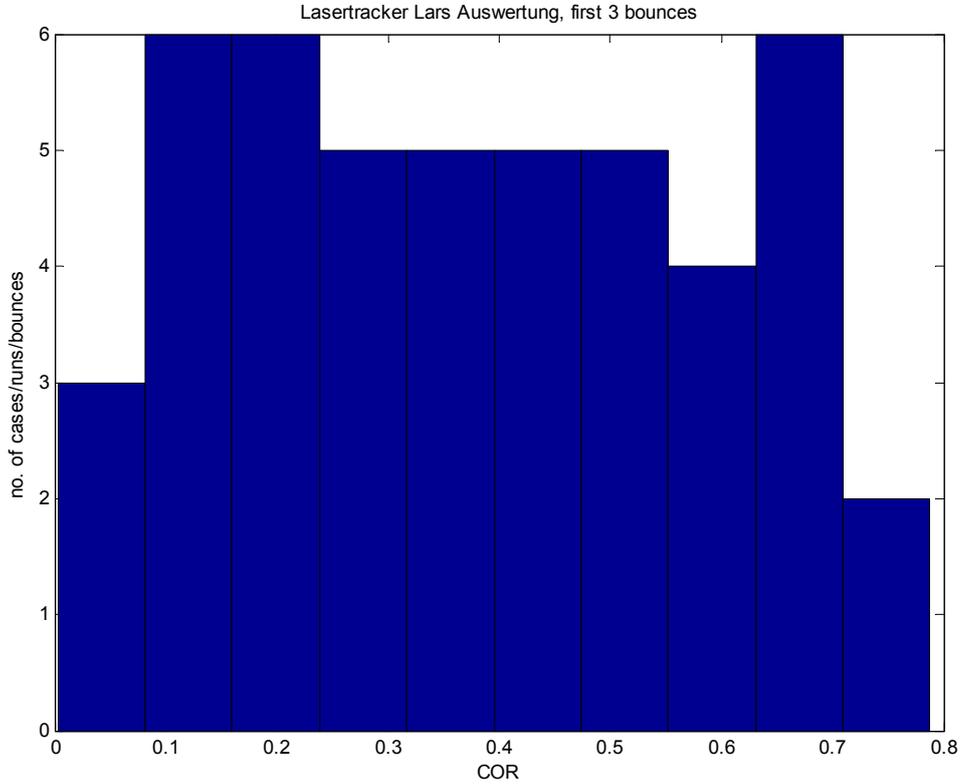

**Figure 14 Histogram of e (Laser data, first 3 bounces only)**

It is obvious that there is not a single COR, but a very wide range from nearly zero to about 0.8 with a median of about 0.4. There is no strong correlation between the attitude at impact and e, except that the cases with the bounce knob show systematically a high e.

Since the uncertainty of e is very variable if the secondary bounces are taken into account (de/e ~ dA/A where A is the amplitude and dA its approximately constant uncertainty), a standard histogram is not the best representation of the results. Assuming that the entirety of e follows a wide probability distribution and each single measured e a normal distribution, we decided to plot the sum of the Gaussian probabilities versus e:

$$p(e) = \sum_i \frac{1}{\sigma_i \sqrt{2\pi}} e^{-\frac{1}{2}\left(\frac{e-e_i}{\sigma_i}\right)^2}$$

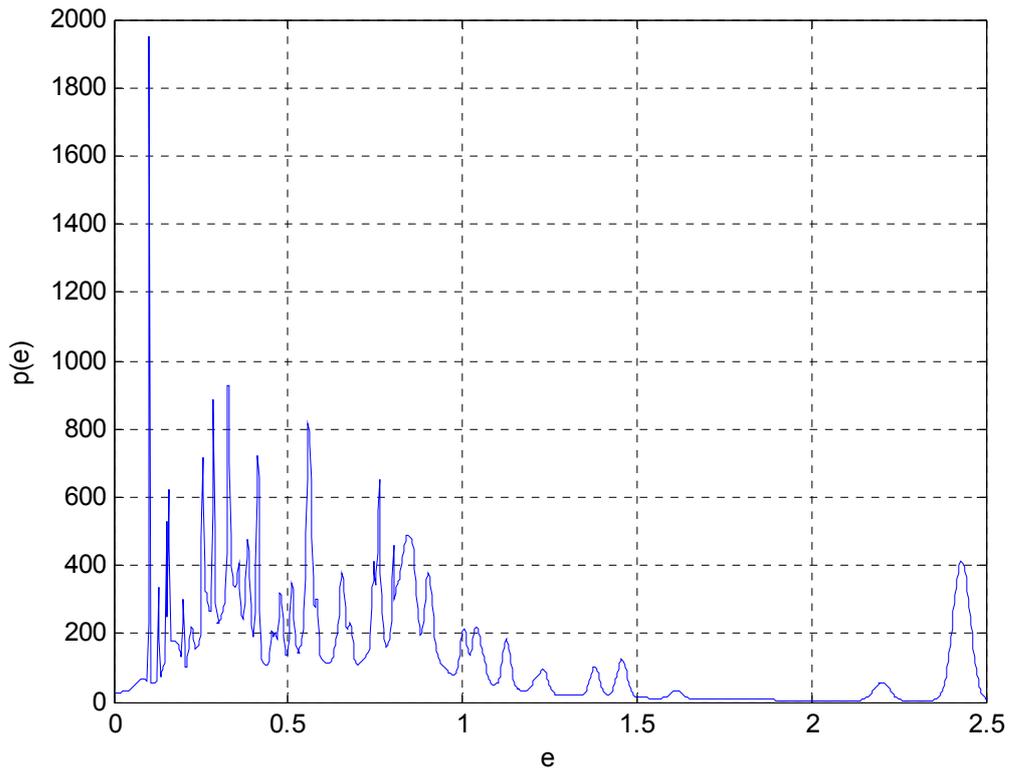

**Figure 15 Laser only, dA provisorisch: chi2red+0.01*A+1 mm.**

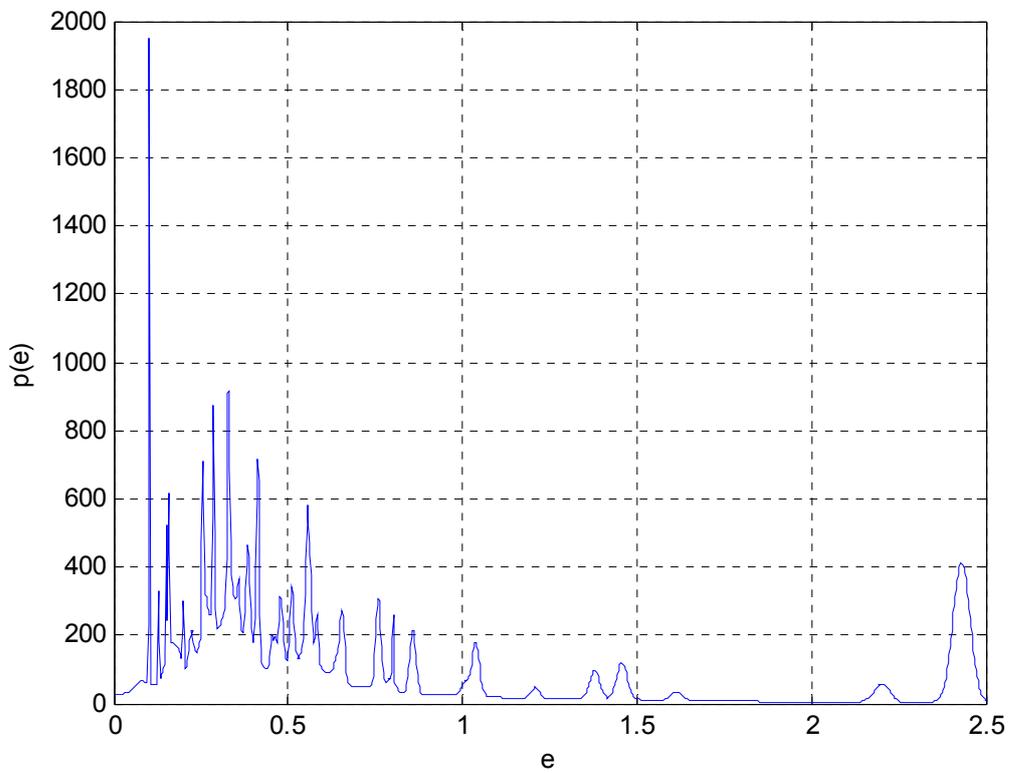

**Figure 16 Same as above but without cases 27,28 (bounce knob)**

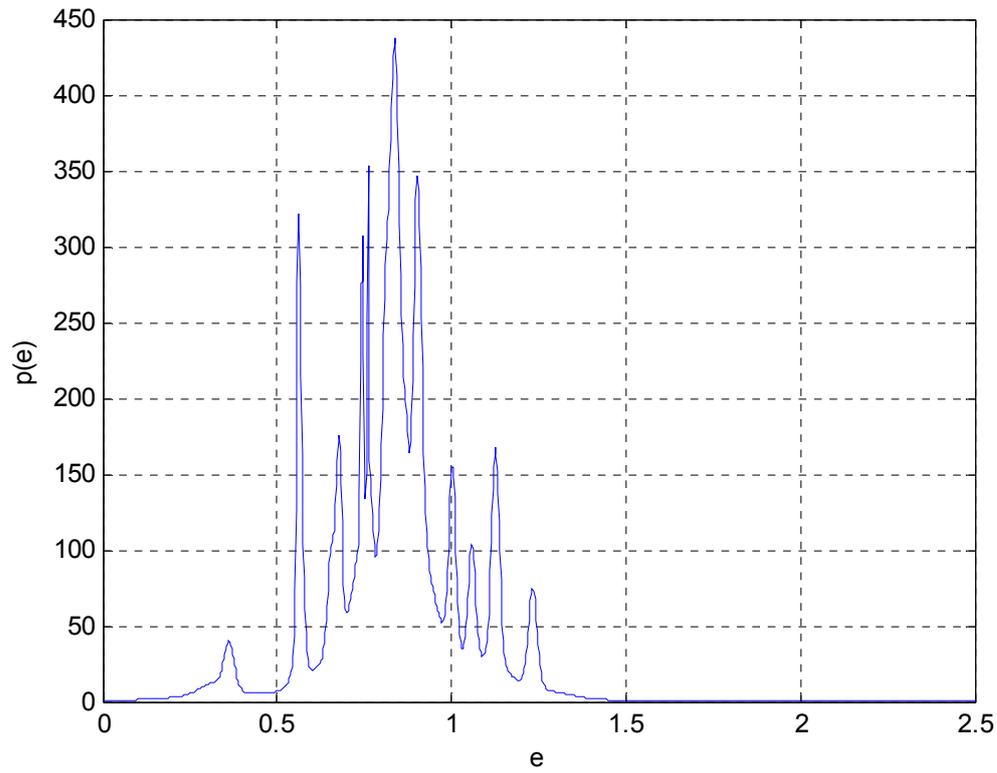

**Figure 17 Only cases 27, 28**

J. Structural COR e*
   TBW

## VIII.  DISCUSSION

The case with the "Bounce knob" is indeed a very special case, we think it is rather unlikely to be realized on an asteroid surface.

## IX.  CONCLUSIONS

Thus the maximum structural COR for MASCOT/Hard wall is ~0.6.

## Acknowledgments

JB wishes to thank the test team (SS, OM, ML, CG) for an excellent work. They conducted the tests with only a rather vague test plan by the PI (JB) and in the absence of the latter, leading to many questions during the analysis which the team patiently answered in great detail. LK was the hard-working intern at DLR who did most of the laborious work of raw data reduction, in particular of the movie data.